\documentclass[reprint,prl,superscriptaddress,amsmath,amssymb,aps,floatfix,nofootinbib]{revtex4-2}
\usepackage[pdftex]{graphicx}
\usepackage{bm}
\usepackage{color}
\usepackage{times}
\usepackage{amsmath}
\usepackage{amssymb}
\usepackage{natbib}
\usepackage{float}

\begin{document}

\title{Fluid-fluid displacement in mixed-wet porous media}

\author{Ashkan Irannezhad}
\affiliation{Department of Civil Engineering, McMaster University, Hamilton, ON, Canada}
\author{Bauyrzhan K. Primkulov}
\affiliation{Department of Civil and Environmental Engineering, Massachusetts Institute of Technology, Cambridge, MA}
\author{Ruben Juanes}
\affiliation{Department of Civil and Environmental Engineering, Massachusetts Institute of Technology, Cambridge, MA}
\author{Benzhong Zhao}
\email{robinzhao@mcmaster.ca}
\affiliation{Department of Civil Engineering, McMaster University, Hamilton, ON, Canada}



\begin{abstract}

It is well-known that wettability exerts fundamental control over multiphase flow in porous media, which has been extensively studied in uniform-wet porous media. In contrast, multiphase flow in porous media with heterogeneous wettability (i.e., mixed-wet) is less well-understood, despite its common occurrence. Here, we study the displacement of silicone oil by water in a mostly oil-wet porous media patterned with discrete water-wet clusters that have precisely controlled wettability. Surprisingly, the macroscopic displacement pattern varies dramatically depending on the details of wettability alteration---the invading water preferentially fills strongly water-wet clusters but encircles weakly water-wet clusters instead, resulting in significant trapping of the defending oil. We explain this counter-intuitive observation with pore-scale simulations, which reveal that the fluid-fluid interfaces at mixed-wet pores resemble an S-shaped saddle with mean curvatures close to zero. We show that incorporation of the capillary entry pressures at mixed-wet pores into a dynamic pore-network model reproduces the experiments. Our work demonstrates the complex nature of wettability control in mixed-wet porous media, and it presents experimental and numerical platforms upon which further insights can be drawn.   

\end{abstract}

\maketitle


Multiphase flow in porous media is of great importance in many natural and industrial settings, including water infiltration~\cite{cueto-prl-2008}, enhanced oil recovery~\cite{orr-science-1984}, geologic carbon sequestration~\cite{szulczewski-pnas-2012}, and electrochemical devices~\cite{zhao-crps-2021}. The displacement of one fluid phase by another in porous media has long been viewed through the lens of Lenormand's diagram, which states that the flow behavior is governed by the viscosity ratio between the fluids, and the relative importance between capillary and viscous forces~\cite{lenormandtouboul88}. Concurrent studies~\cite{stokes-prl-1986,cieplak-prl-1988} demonstrated that the fluids’ relative affinity to the porous media (i.e., wettability) also has a profound influence on the flow behavior. Specifically, the displacement of a less wetting fluid by a more wetting fluid (i.e., imbibition) yields more compact displacement patterns compared to the displacement of a more wetting fluid by a less wetting fluid (i.e., drainage). More recently, a systematic study of fluid-fluid displacement in microfluidics~\cite{zhao-pnas-2016} illustrated wettability control over a wide range of wettability conditions, which culminated in the extension of Lenormand's diagram to include wettability~\cite{primkulov-jfm-2021}. A common theme between these studies is the spatial uniformity of the porous media's wettability. In contrast, multiphase flow in porous media with spatially heterogeneous wettability (i.e., mixed-wet) condition is less well-understood.     

Mixed-wet porous media is a common occurence in many settings~---~for example, parts of a groundwater aquifer become water-repellent after contacting a non-aqueous phase liquid (NAPL)~\cite{powers-jee-1996}, and portions of a gas diffusion layer become water-repellent after the addition of polytetrafluoroethylene (PTFE)~\cite{sinha-ces-2008}. Mixed-wettability is particularly prevalent in oil-bearing reservoir rocks~\cite{Salathiel1973,kovscek-aiche-1993}. High-resolution in situ wettability characterizations of reservoir rocks demonstrated the presence of a wide range of contact angle within a single millimeter-sized core sample, and the existence of distinct water-wet and water-repellent regions~\cite{andrew2014pore,AlRatrout2018,blunt-jcis-2019}. 

\begin{figure*}[htp]
	\centering
	\includegraphics[width=17.2cm]{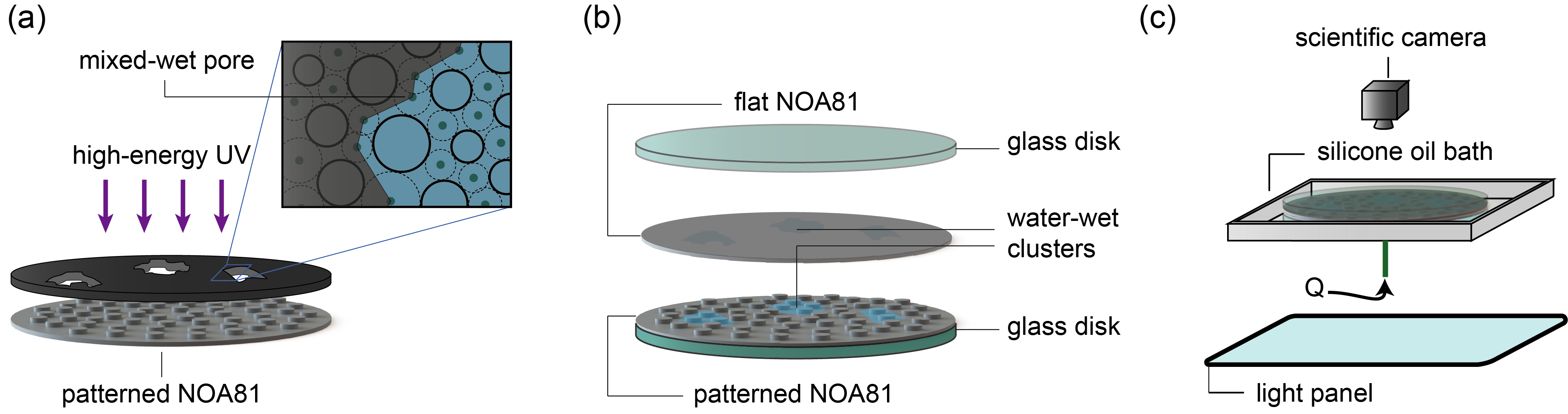}
	\caption{We conduct radial fluid-fluid displacement experiments in microfluidic flow cells with spatially heterogeneous wettabilities. (a) The bottom half of the flow cell is patterned with $\sim{16,000}$ cylindrical posts of height $b=160~\mu$m. The flow cell is made of a photocurable polymer (NOA81) that is hydrophobic in nature, but becomes more hydrophilic after exposure to high-energy UV radiation. We cover the NOA81 surface with a photomask during UV exposure to achieve water-wet clusters. The border between oil-wet and water-wet regions are delineated by connecting the centers of mixed-wet pores (inset). (b) The top half of the flow cell consists of a flat sheet of NOA81 patterned with the same spatially heterogeneous wettability and it is precision aligned with the bottom half. (c) The microfluidic cell is initially saturated with a viscous silicone oil ($\eta_{\rm{oil}}=50$~mPa$\cdot$s) and placed in a bath of the same fluid to avoid capillary edge effects. We inject water into the center of the microfluidic cell at a constant rate $Q$ and image the experiment from above with a scientific camera.}\label{fig:setup}
\end{figure*}

Recent experiments in mixed-wet reservoir rocks demonstrated significant differences in flow behavior compared to their homogeneously-wet counterparts. For instance, imbibition in oil-saturated mixed-wet core sample displaced more oil~\cite{lin-pre-2019,alhammadi-fuel-2020,scanziani-royal-2020}, while imbibition in CO$_2$-saturated mixed-wet core sample resulted in less residual trapping~\cite{almenhali-krevor-est-2016,almenhali-est-2016}. Additional features of multiphase flow in mixed-wet core samples include increases in dynamic ganglion flow~\cite{zou-wrr-2018,rucker-grl-2019} and the presence of fluid-fluid interfaces with very low mean curvature~\cite{lin-pre-2019,scanziani-royal-2020}.

Here, we use patterned microfluidic flow cells to study viscously unfavorable displacement of silicone oil by water in mixed-wet porous media. This system allows for simultaneous visualization of the macroscopic displacement front and the pore-scale fluid-fluid interface. Additionally, the flow cells are designed with precise pore geometries and mixed-wettabilities in the form of discrete water-wet clusters, which eliminate the uncertainty associated with natural media and enable direct mapping of observations to the interfacial fluid dynamics at the pore scale. Our results show that the macroscopic displacement pattern is sensitive to the exact wettability of the water-wet clusters---while the invading water preferentially fills \emph{strongly water-wet clusters}, it encircles \emph{weakly water-wet clusters} instead and traps a significant amount of the defending oil. We show that this dramatic difference is caused by the unique morphology of the fluid-fluid interface in mixed-wet pores, which resembles an S-shaped saddle with mean curvature close to zero. The pore-scale fluid-fluid interfaces dictate the capillary entry pressures, which lead to the preferential filling of mixed-wet pores over \emph{weakly water-wet pores}, but not \emph{strongly water-wet pores}. We incorporate the pore-scale insights into a dynamic pore-network model, which reproduces the experiments across different flow rates and wettability conditions.


\emph{Experiments in Mixed-wet Microfluidics.}---We conduct fluid-fluid displacement experiments in microfluidic flow cells with spatially heterogeneous wettability conditions (i.e. mixed-wet). Each flow cell contains $\sim$16,000 cylindrical posts and it is fabricated with a photocurable polymer (NOA81, Norland Optical Adhesives). The NOA81 surface is oil-wet in nature, but it becomes increasingly water-wet with exposure to high-energy UV irradiation~\cite{levache-labchip-2012,levache-prl-2014,zhao-pnas-2016,Odier2017}. We expose the flow cell locally to high-energy UV via the application of a UV-blocking mask. The mask is patterned with cutouts that yields four distinct water-wet clusters~(Fig.~\ref{fig:setup}a). We characterize the wettability of the flow cell using the static advancing contact angle $\theta$ of water immersed in silicone oil. The bulk of the flow cell is oil-wet ($\theta=120^{\circ}$), while the clusters are either weakly water-wet ($\theta=60^{\circ}$) or strongly water-wet ($\theta=30^{\circ}$). We fabricate a new flow cell for each experiment to ensure precise control over its wettability~(Fig.~\ref{fig:setup}b).

To perform an experiment, we first saturate the flow cell with a viscous silicone oil ($\eta_{\rm{oil}}=50$~mPa$\cdot$s). We then inject deionized water ($\eta_{\rm{water}}=0.99$~mPa$\cdot$s) into the center of the flow cell at a constant volumetric rate $Q$ to displace the ambient silicone oil~(Fig.~\ref{fig:setup}c). This is a viscously unfavorable displacement, with viscosity ratio $\mathcal{M} =\eta_\mathrm{oil}/\eta_\mathrm{water} \approx{50}$. We characterize the relative importance between viscous forces and capillary forces using the macroscopic capillary number $\rm{Ca}=\eta_{\rm{oil}}v_{inj}/\gamma$~\cite{lenormandtouboul88}, where $\gamma=13\pm2$~mN/m is the interfacial tension between the fluids and $\mathrm{v_{inj}}=Q/(bd)$ is the characteristic injection velocity as constrained by the gap thickness~$b$ and the median pore-throat size~$d$. We conduct experiments at varying injection rates ($Q=0.0003$, $0.03$ and $0.3$~mL/min), which correspond to capillary numbers spanning three orders of magnitude~($\rm{Ca}=6\times{10^{-4}}$, $6\times{10^{-2}}$, and $6\times{10^{-1}}$, respectively) for both mixed-wet flow cells with weakly water-wet clusters and flow cells with strongly water-wet clusters. We provide a detailed description of the flow-cell fabrication process and the experimental procedure in the \textit{Supplemental Material}~\cite{supplement}.   

\begin{figure*}[htp]
	\centering
	\includegraphics[width=12.9cm]{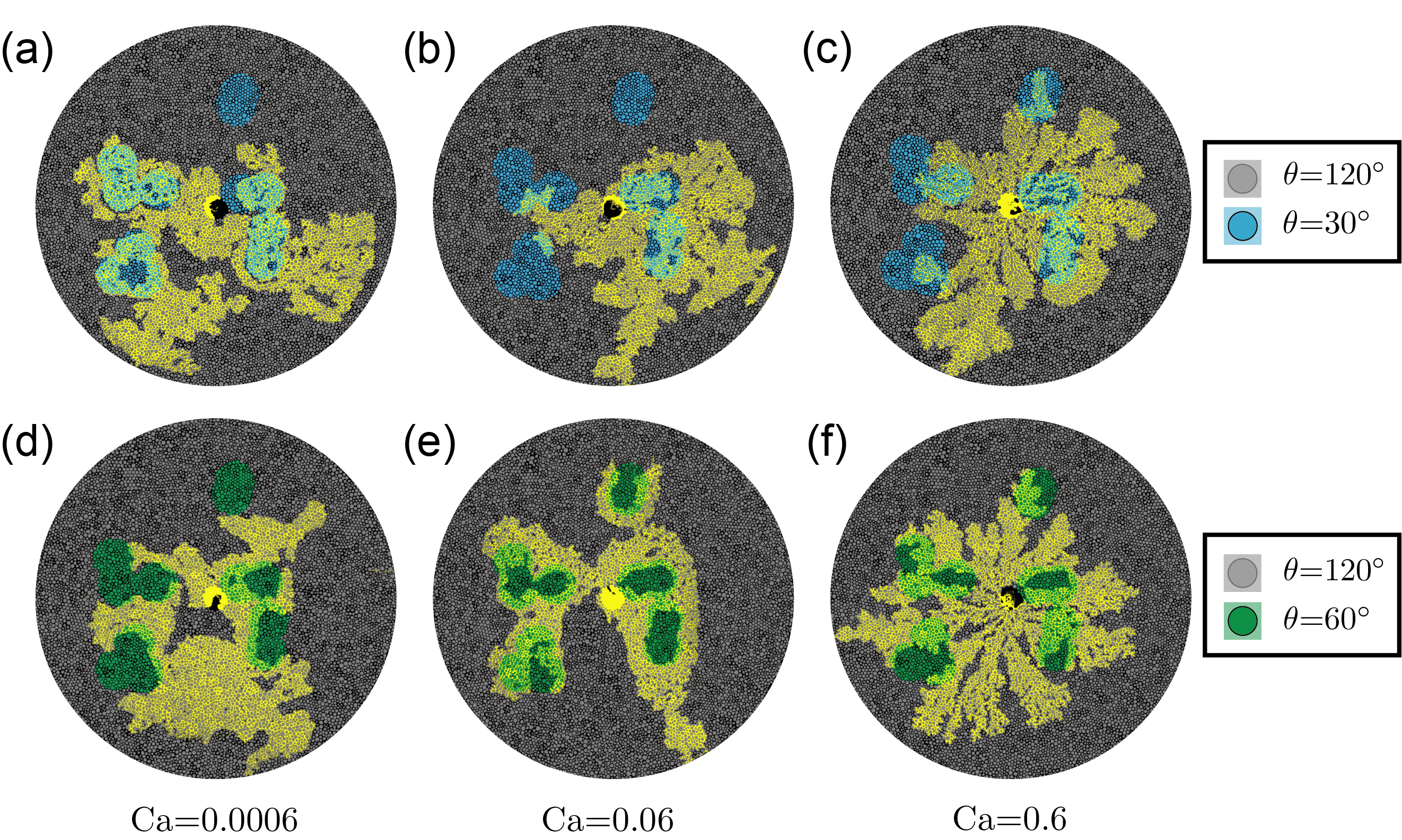}
	\caption{\emph{Experimental} displacement patterns of water (yellow) displacing silicone oil (black) in mixed-wet microfluidic flow cells at three distinct capillary numbers. The bulk of the flow cell is oil-wet (gray posts, $\theta=120^{\circ}$), and it is interspersed either with strongly water-wet clusters (blue posts, $\theta=30^{\circ}$), or with  weakly water-wet clusters (green posts, $\theta=60^{\circ}$). The patterns correspond to when the invading water reaches the perimeter of the flow cell and they are oriented in the same way to aid visual comparison. The invading water saturates the strongly water-wet clusters (top row), but encircles the weakly water-wet clusters (bottom row).}
	\label{fig:experimental_results}
\end{figure*}


Intuitively, one would expect that the invading water preferentially fills the water-wet clusters in its path, especially at low Ca where capillary forces dominate. This is indeed the case for experiments in the mixed-wet flow cells with \emph{strongly water-wet clusters}~(Fig.~\ref{fig:experimental_results}a--c and Movie S1). Surprisingly, we observe the opposite behavior in the mixed-wet flow cells with \emph{weakly water-wet clusters}, where the water advances by encircling the weakly water-wet clusters instead of saturating them~(Fig.~\ref{fig:experimental_results}d--f and Movie S2). 

\begin{figure}[htp]
	\centering
	\includegraphics[width=8.6cm]{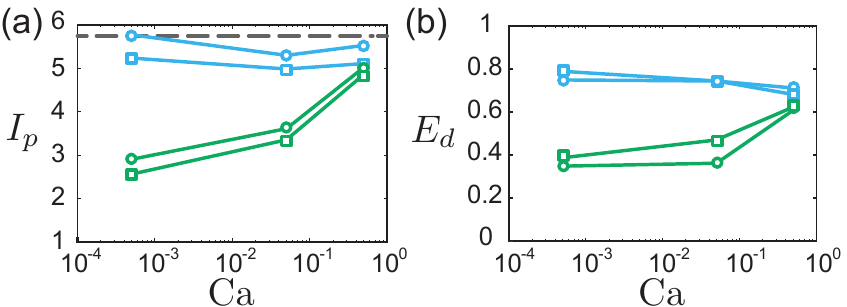}
	\caption{(a) Water-wet pore preference index $I_\textrm{p}$ as a function of Ca, where $I_\textrm{p}$ is defined as the ratio between the number of invaded water-wet pore throats and the number of invaded mixed-wet pore throats. Experimental $I_\textrm{p}$ for flow cells with weakly water-wet clusters (green circles) and strongly water-wet clusters (blue circles) are plotted alongside numerical $I_\textrm{p}$ (squares) obtained from our pore-network model (see Fig.~\ref{fig:numerical_results}). The gray dashed line represents the ratio between the total number of water-wet pore throats and the total number of mixed-wet pore throats in the entire flow cell. (b) Water-wet cluster displacement efficiency $E_\textrm{d}$ as a function of Ca, where $E_\textrm{d}$ is defined as the fraction of the defending fluid displaced from the water-wet cluster closest to the injection port. $E_\textrm{d}$ for flow cells with weakly water-wet clusters (green circles) and strongly water-wet clusters (blue circles) are plotted alongside numerical $E_\textrm{d}$ (squares) obtained from our pore-network model (see Fig.~\ref{fig:numerical_results}).}
	\label{fig:plots}
\end{figure}

To provide quantitative insight into fluid-fluid displacement through the water-wet clusters, we calculate two metrics for each experiment: (i) The water-wet pore preference index $I_\textrm{p}$, which is the ratio of the number of invaded water-wet pore throats to the number of invaded mixed-wet pore throats at the end of the experiment~(Fig.~\ref{fig:plots}a). We define a mixed-wet pore throat as one that is between an oil-wet post and a water-wet post (Fig.~\ref{fig:setup}a); (ii) The water-wet cluster displacement efficiency $E_\textrm{d}$, which is defined as the fraction of the defending fluid displaced from the water-wet cluster closest to the injection port at the end of the experiment~(Fig.~\ref{fig:plots}b). We confine $E_\textrm{d}$ measurements to the water-wet cluster closest to the injection port because not all water-wet clusters come in the path of the invading water during the experiment.

We calculate the ratio between the total number of water-wet pore throats and the total number of mixed-wet pore throats in the entire flow cell as the upper bound for $I_p$ (i.e., $I_p^{\text{max}}=5.8$), which correspond to the case where all encountered water-wet clusters are completely saturated by the invading water. We find high water-wet pore preference index ($I_p>5$) in the flow cell with strongly water-wet clusters at all Ca. Decreasing the affinity of the water-wet clusters to the invading fluid (i.e., increasing $\theta$ from $30^\circ$ to $60^\circ$) dramatically decreases $I_p$. This is especially distinct at low Ca ($I_p<3$), where the invading water enters the mixed-wet pore throats along the perimeter of the water-wet clusters without saturating them (Fig.~\ref{fig:experimental_results}d--e). However, $I_p$ in the flow cell with weakly water-wet clusters does increase at high Ca, when viscous forces become more important compared to capillary forces~(Fig.~\ref{fig:plots}a). Similar observations have been reported by Armstrong \& Wildenschild~\citep{Armstrong2012}, whose X-ray computed microtomography (micro-CT) imaging of water flooding in oil-saturated mixed-wet cores showed that fluid-fluid interfaces are preferentially located in the mixed-wet pores. 

At low Ca, the displacement efficiency is much higher in the flow cell with strongly water-wet clusters ($E_d=0.78$) compared to the flow cell with weakly water-wet clusters ($E_d=0.38$), since the preferential filling of mixed-wet pores along the perimeter of the weakly water-wet clusters traps a significant amount of the defending oil. As Ca increases, $E_d$ decreases in the flow cell with strongly water-wet clusters as a result of viscous fingering. In contrast, $E_d$ increases in the flow cell with weakly water-wet clusters. We find similar $E_d$ values between the two mixed-wet flow cells at the highest Ca, where viscous forces dominate~(Fig.~\ref{fig:plots}b).  

\begin{figure}[ht]
	\centering
	\includegraphics[width=8.6cm]{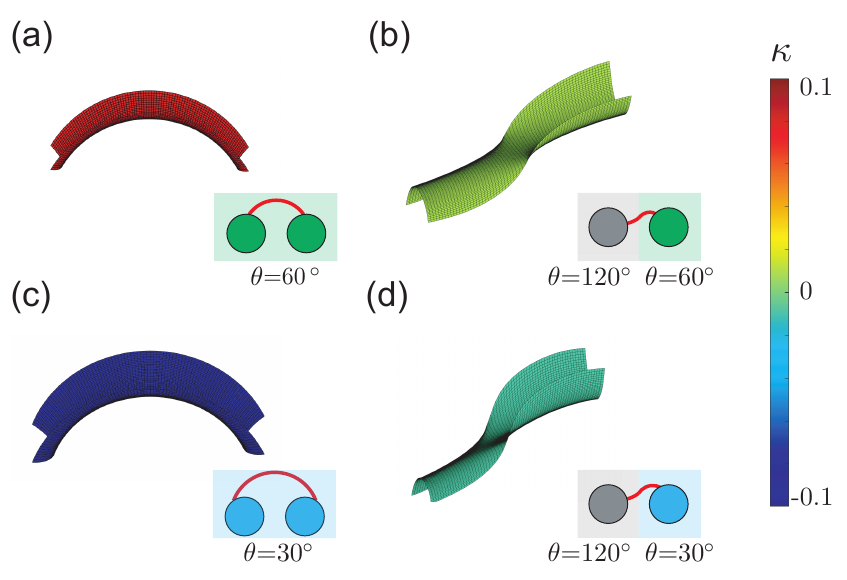}
	\caption{Three-dimensional visualization of the last stable fluid-fluid interface through (a) a uniform-wet pore consisting of a pair of weakly water-wet posts, (b) a mixed-wet pore consisting of an oil-wet post next to a weakly water-wet post, (c) a uniform-wet pore consisting of a pair of strongly water-wet posts, and (d) a mixed-wet pore consisting of an oil-wet post next to a strongly water-wet post. The colormap shows the local mean curvature of the meniscus. The last stable fluid-fluid interface through a uniform-wet pore consisting of \emph{weakly water-wet} posts has a \emph{higher} mean curvature compared to its mixed-wet counterpart. However, the last stable fluid-fluid interface through a uniform-wet pore consisting of \emph{strongly water-wet} posts has a \emph{lower} mean curvature compared to its mixed-wet counterpart. The differences in interface curvature dictate invasion sequence and lead to drastically different displacement patterns (Fig.~2).}
	\label{fig:SE}
\end{figure}

\emph{Pore-scale Physics.}---The significant $I_\textrm{p}$ and $E_\textrm{d}$ differences between flow cells with strongly water-wet clusters and those with weakly water-wet clusters at low Ca indicate that capillarity at mixed-wet pores play a fundamental role. To gain a deeper understanding of wettability control at the pore-scale, we investigate the fluid-fluid interface using \texttt{Surface Evolver}, which is a finite element solver that minimizes the overall surface energy of a fluid-fluid-solid system~\cite{brakke1992surface}. Specifically, we simulate the quasi-static evolution of the fluid-fluid interface in 3D through a typical uniform-wet pore throat versus a typical mixed-wet pore throat. Mixed-wet pore throats exist between a water-wet post and an oil-wet post, and we assign the wettability of the top and bottom surfaces such that the boundary between water-wet and oil-wet regions bisect the pore throat, which is similar to the experimental configuration~(Fig.~\ref{fig:setup}a inset). We report the 3D shape of the last stable fluid-fluid interface, which occurs at the point beyond which no viable solution is possible~(Fig.~\ref{fig:SE}). 
Due to the equilibrium condition, each fluid-fluid interface has constant mean curvature $\kappa$ in space.

\begin{figure}[ht]
	\centering
	\includegraphics[width=8.6cm]{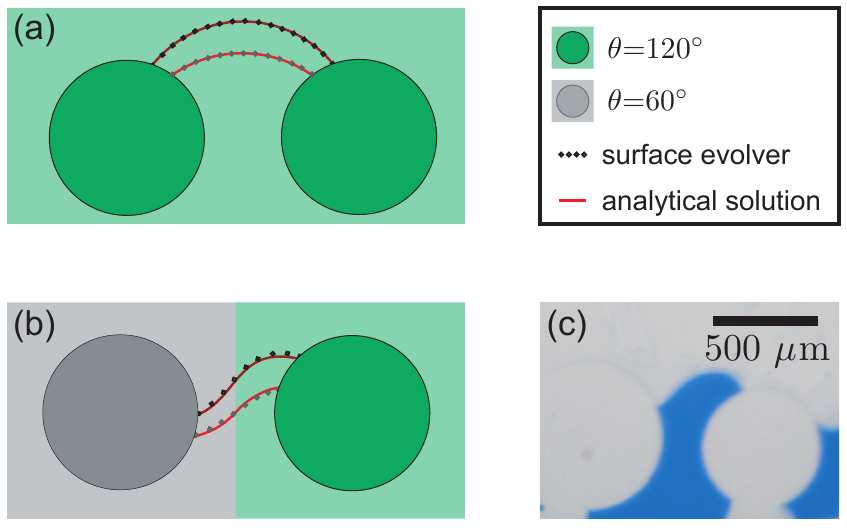}
	\caption{In-plane curvature of the fluid-fluid interface at (a) a typical uniform-wet pore and (b) a typical mixed-wet pore. The circles represent simulation results obtained via \texttt{Surface Evolver}, whereas solid lines represent our analytical solutions~\cite{supplement}. (c) Experimental snapshot of the meniscus at a mixed-wet pore consisting of an oil-wet post next to a weakly water-wet post at $\text{Ca}=6\times{10^{-4}}$.}
	\label{fig:SE_vs_analytic}
\end{figure}

The morphology of fluid-fluid interfaces at mixed-wet pore throats are distinct from those at uniform-wet pore throats. Specifically, the fluid-fluid interfaces at mixed-wet pore throats resemble an S-shaped saddle with mean curvatures close to zero~(Fig.~\ref{fig:SE}b, d). Pore-scale imaging of our experiments indeed shows the existence of S-shaped fluid-fluid interfaces at mixed-wet pore throats~(Fig.~\ref{fig:SE_vs_analytic}c). We note that saddle-shaped fluid-fluid interfaces with zero mean curvature have recently been observed in X-ray micro-CT imaging of waterflooding in oil-saturated Bentheimer sandstone cores~\cite{lin-pre-2019}. For the flow cell with \emph{weakly water-wet clusters}, the mean curvature of the fluid-fluid interface at a mixed-wet pore throat is lower than that at a uniform-wet pore throat (Fig.~\ref{fig:SE}a, b). Since capillary entry pressure is given by $P_c=\gamma\kappa$, the lower mean curvature is responsible for the preferential filling of mixed-wet pore throats in weakly water-wet clusters. For the flow cell with \emph{strongly water-wet clusters}, the mean curvature of the fluid-fluid interface at a mixed-wet pore throat is higher than that at a uniform-wet pore throat (Fig.~\ref{fig:SE}c, d), which is responsible for the preferential filling of water-wet pores.

\emph{Numerical Simulation of Fluid-fluid Displacement in Mixed-Wet Porous Media.}---There are several classes of computational approaches to simulate pore-scale fluid-fluid displacement in porous media, including lattice/particle-based methods (e.g., lattice Boltzmann method), upscaled continuum methods (e.g., phase-field model), and topological methods (e.g., pore-network model)~\cite{zhao-pnas-2019}. Pore-network models stand out in their relative simplicity and low computational demand~\cite{blunt-cocis-2001}. Additionally, the pioneering work of Cieplak and Robbins demonstrated that pore network models with quasistatic interface tracking can capture the impact of wettability on the macroscopic displacement pattern~\cite{cieplak-prl-1988,cieplakrobbins90}. This quasistatic framework was later extended to include the dynamics of flow by coupling the critical capillary pressure with viscous pressure drop at the pore scale~\cite{holtzman-prl-2015,Primkulov2019,primkulov-jfm-2021}.

To capture the pore-scale physics in our experiments, we derive analytical expressions for the fluid-fluid interface evolution through both uniform-wet and mixed-wet pore throats (Fig.~\ref{fig:SE_vs_analytic}; Fig.~S4~\cite{supplement}). This allows us to calculate the critical capillary pressure at each pore throat along the invasion front, which takes place when the fluid-fluid interface encounters a burst, touch or overlap event. The burst event is equivalent to a Haines jump~\cite{berg-pnas-2013,haines-jas-1930}, while the touch event occurs when the fluid-fluid interface touches the closest opposing post, and the overlap event occurs when the fluid-fluid interface merges with a neighboring interface~\cite{cieplak-prl-1988,cieplakrobbins90} (Fig.~S5~\cite{supplement}). Finally, we incorporate the critical capillary pressure in the dynamic pore-network model framework of Primkulov et al.~\cite{primkulov-jfm-2021} to arrive at a pore-scale model of fluid-fluid displacement in mixed-wet porous media.

\begin{figure*}[htp]
	\centering
	\includegraphics[width=12.9cm]{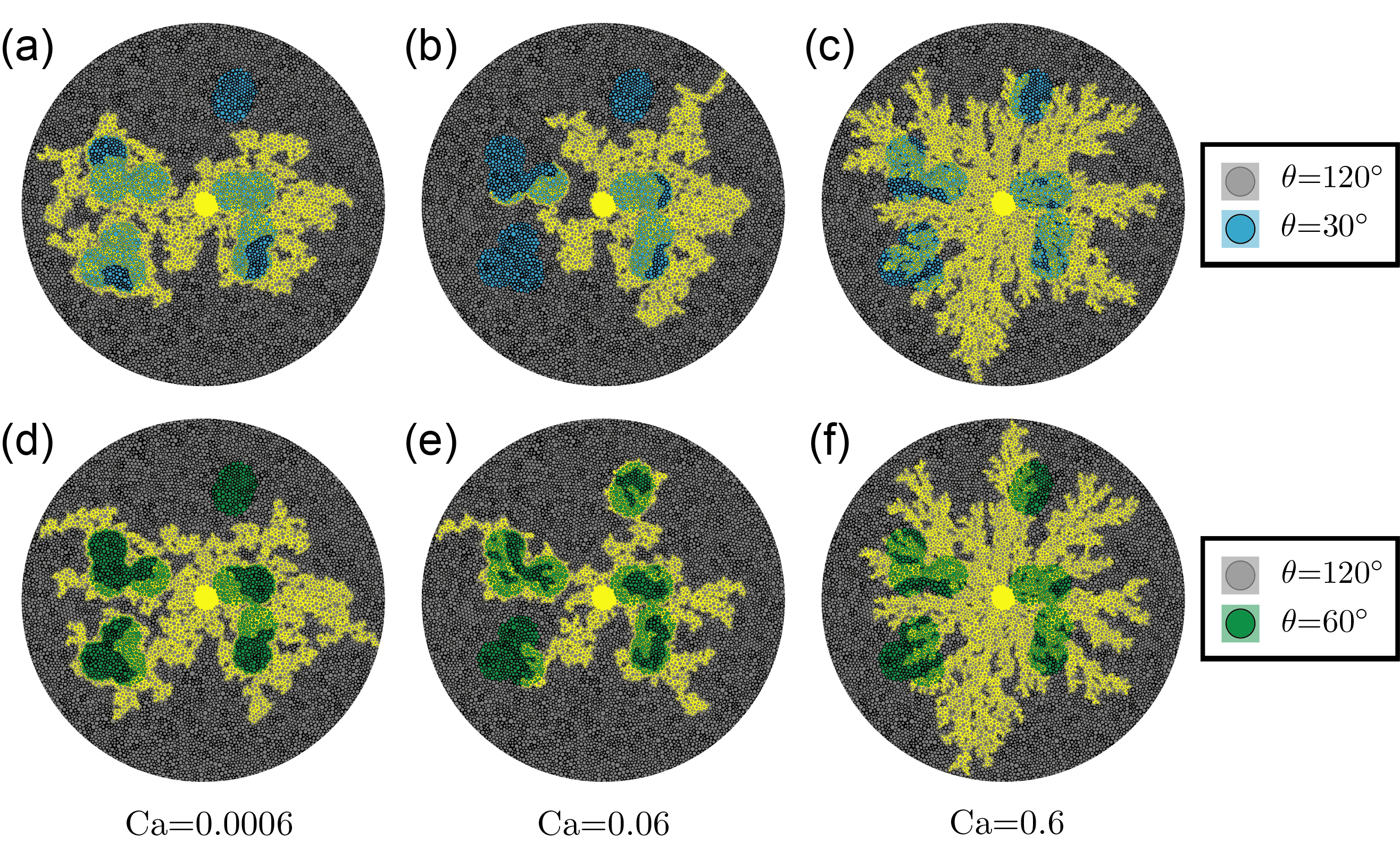}
	\caption{\emph{Numerical} displacement patterns simulated by the dynamic pore network model for the same conditions as the experiments. The simulations closely resemble the experiments (Fig. 2), and they capture the contrasting behaviors of water invasion in clusters of weakly water-wet posts versus clusters of strongly water-wet posts.}
	\label{fig:numerical_results}
\end{figure*}

We apply the pore-network model to simulate the constant rate displacement of silicone oil by water in our experiments (Fig.~\ref{fig:numerical_results}). Qualitatively, the simulations capture the salient features of the experiments---the invading water preferentially fills the \emph{strongly water-wet clusters} but encircles the \emph{weakly water-wet clusters} at low Ca. Quantitatively, the water-wet pore preference index $I_p$ and water-wet cluster displacement efficiency $E_d$ extracted from the simulations reproduce the experimental measurements across all Ca (Fig.~\ref{fig:plots}). 

We have developed a unique experimental platform to study multiphase flow in mixed-wet porous media via the displacement of silicone oil by water in predominantly oil-wet microfluidic flow cells patterned with well-controlled water-wet clusters. Our results show the macroscopic fluid-fluid displacement pattern is highly sensitive to the the details of wettability alteration---while the invading water preferentially fills \emph{strongly water-wet clusters}, it encircles \emph{weakly water-wet clusters} instead and traps a significant amount of the defending oil (Fig.~\ref{fig:experimental_results}). This surprising finding stems from the unique morphology of the fluid-fluid interface in mixed-wet pores, which resembles an S-shaped saddle with mean curvature close to zero (Fig.~\ref{fig:SE}). As a result, the critical capillary pressure of a typical mixed-wet pore is lower than that of a \emph{weakly water-wet pore}, but higher than that of a \emph{strongly water-wet pore}. We show that the pore-scale fluid-fluid interface at mixed-wet pores can be captured by simple analytical expressions (Fig.~\ref{fig:SE_vs_analytic}) and incorporated in a dynamic pore-network model, which reproduces our experimental observations across different wettability conditions and Ca (Fig.~\ref{fig:numerical_results}).

Our results highlight the nuanced, yet critical role wettability plays in multiphase flow in mixed-wet porous media. Meanwhile, the experimental and numerical platforms introduced here provide a controlled complement to traditional core-flooding experiments in natural porous media. These tools enable investigation of the many parameters that impact multiphase flow in mixed-wet porous media, including wettability contrast between clusters of different wetting properties, cluster size and spatial distribution to name a few.

\bibliography{mixwet}

\begin{thebibliography}{34}%
\makeatletter
\providecommand \@ifxundefined [1]{%
 \@ifx{#1\undefined}
}%
\providecommand \@ifnum [1]{%
 \ifnum #1\expandafter \@firstoftwo
 \else \expandafter \@secondoftwo
 \fi
}%
\providecommand \@ifx [1]{%
 \ifx #1\expandafter \@firstoftwo
 \else \expandafter \@secondoftwo
 \fi
}%
\providecommand \natexlab [1]{#1}%
\providecommand \enquote  [1]{``#1''}%
\providecommand \bibnamefont  [1]{#1}%
\providecommand \bibfnamefont [1]{#1}%
\providecommand \citenamefont [1]{#1}%
\providecommand \href@noop [0]{\@secondoftwo}%
\providecommand \href [0]{\begingroup \@sanitize@url \@href}%
\providecommand \@href[1]{\@@startlink{#1}\@@href}%
\providecommand \@@href[1]{\endgroup#1\@@endlink}%
\providecommand \@sanitize@url [0]{\catcode `\\12\catcode `\$12\catcode
  `\&12\catcode `\#12\catcode `\^12\catcode `\_12\catcode `\%12\relax}%
\providecommand \@@startlink[1]{}%
\providecommand \@@endlink[0]{}%
\providecommand \url  [0]{\begingroup\@sanitize@url \@url }%
\providecommand \@url [1]{\endgroup\@href {#1}{\urlprefix }}%
\providecommand \urlprefix  [0]{URL }%
\providecommand \Eprint [0]{\href }%
\providecommand \doibase [0]{https://doi.org/}%
\providecommand \selectlanguage [0]{\@gobble}%
\providecommand \bibinfo  [0]{\@secondoftwo}%
\providecommand \bibfield  [0]{\@secondoftwo}%
\providecommand \translation [1]{[#1]}%
\providecommand \BibitemOpen [0]{}%
\providecommand \bibitemStop [0]{}%
\providecommand \bibitemNoStop [0]{.\EOS\space}%
\providecommand \EOS [0]{\spacefactor3000\relax}%
\providecommand \BibitemShut  [1]{\csname bibitem#1\endcsname}%
\let\auto@bib@innerbib\@empty
\bibitem [{\citenamefont {Cueto-Felgueroso}\ and\ \citenamefont
  {Juanes}(2008)}]{cueto-prl-2008}%
  \BibitemOpen
  \bibfield  {author} {\bibinfo {author} {\bibfnamefont {L.}~\bibnamefont
  {Cueto-Felgueroso}}\ and\ \bibinfo {author} {\bibfnamefont {R.}~\bibnamefont
  {Juanes}},\ }\bibfield  {title} {\bibinfo {title} {Nonlocal interface
  dynamics and pattern formation in gravity-driven unsaturated flow through
  porous media},\ }\href@noop {} {\bibfield  {journal} {\bibinfo  {journal}
  {Phys. Rev. Lett.}\ }\textbf {\bibinfo {volume} {101}},\ \bibinfo {pages}
  {244504} (\bibinfo {year} {2008})}\BibitemShut {NoStop}%
\bibitem [{\citenamefont {Orr}\ and\ \citenamefont
  {Taber}(1984)}]{orr-science-1984}%
  \BibitemOpen
  \bibfield  {author} {\bibinfo {author} {\bibfnamefont {F.}~\bibnamefont
  {Orr}}\ and\ \bibinfo {author} {\bibfnamefont {J.}~\bibnamefont {Taber}},\
  }\bibfield  {title} {\bibinfo {title} {Use of carbon dioxide in enhanced oil
  recovery},\ }\href@noop {} {\bibfield  {journal} {\bibinfo  {journal}
  {Science}\ }\textbf {\bibinfo {volume} {224}},\ \bibinfo {pages} {563}
  (\bibinfo {year} {1984})}\BibitemShut {NoStop}%
\bibitem [{\citenamefont {Szulczewski}\ \emph {et~al.}(2012)\citenamefont
  {Szulczewski}, \citenamefont {MacMinn}, \citenamefont {Herzog},\ and\
  \citenamefont {Juanes}}]{szulczewski-pnas-2012}%
  \BibitemOpen
  \bibfield  {author} {\bibinfo {author} {\bibfnamefont {M.~L.}\ \bibnamefont
  {Szulczewski}}, \bibinfo {author} {\bibfnamefont {C.~W.}\ \bibnamefont
  {MacMinn}}, \bibinfo {author} {\bibfnamefont {H.~J.}\ \bibnamefont
  {Herzog}},\ and\ \bibinfo {author} {\bibfnamefont {R.}~\bibnamefont
  {Juanes}},\ }\bibfield  {title} {\bibinfo {title} {Lifetime of carbon capture
  and storage as a climate-change mitigation technology},\ }\href@noop {}
  {\bibfield  {journal} {\bibinfo  {journal} {Proc. Natl. Acad. Sci. USA}\
  }\textbf {\bibinfo {volume} {109}},\ \bibinfo {pages} {5185} (\bibinfo {year}
  {2012})}\BibitemShut {NoStop}%
\bibitem [{\citenamefont {Zhao}\ \emph {et~al.}(2021)\citenamefont {Zhao},
  \citenamefont {Lee}, \citenamefont {Lee}, \citenamefont {Fahy}, \citenamefont
  {La{M}anna}, \citenamefont {Baltic}, \citenamefont {Jacobson}, \citenamefont
  {Hussey},\ and\ \citenamefont {Bazylak}}]{zhao-crps-2021}%
  \BibitemOpen
  \bibfield  {author} {\bibinfo {author} {\bibfnamefont {B.}~\bibnamefont
  {Zhao}}, \bibinfo {author} {\bibfnamefont {C.~H.}\ \bibnamefont {Lee}},
  \bibinfo {author} {\bibfnamefont {J.~K.}\ \bibnamefont {Lee}}, \bibinfo
  {author} {\bibfnamefont {K.~F.}\ \bibnamefont {Fahy}}, \bibinfo {author}
  {\bibfnamefont {J.~M.}\ \bibnamefont {La{M}anna}}, \bibinfo {author}
  {\bibfnamefont {E.}~\bibnamefont {Baltic}}, \bibinfo {author} {\bibfnamefont
  {D.~L.}\ \bibnamefont {Jacobson}}, \bibinfo {author} {\bibfnamefont {D.~S.}\
  \bibnamefont {Hussey}},\ and\ \bibinfo {author} {\bibfnamefont
  {A.}~\bibnamefont {Bazylak}},\ }\bibfield  {title} {\bibinfo {title}
  {Superhydrophilic porous transport layer enhances efficiency of polymer
  electrolyte membrane electrolyzers},\ }\href@noop {} {\bibfield  {journal}
  {\bibinfo  {journal} {Cell Rep. Physical Science}\ }\textbf {\bibinfo
  {volume} {2}},\ \bibinfo {pages} {100580} (\bibinfo {year}
  {2021})}\BibitemShut {NoStop}%
\bibitem [{\citenamefont {Lenormand}\ \emph {et~al.}(1988)\citenamefont
  {Lenormand}, \citenamefont {Touboul},\ and\ \citenamefont
  {Zarcone}}]{lenormandtouboul88}%
  \BibitemOpen
  \bibfield  {author} {\bibinfo {author} {\bibfnamefont {R.}~\bibnamefont
  {Lenormand}}, \bibinfo {author} {\bibfnamefont {E.}~\bibnamefont {Touboul}},\
  and\ \bibinfo {author} {\bibfnamefont {C.}~\bibnamefont {Zarcone}},\
  }\bibfield  {title} {\bibinfo {title} {Numerical models and experiments on
  immiscible displacements in porous media},\ }\href@noop {} {\bibfield
  {journal} {\bibinfo  {journal} {J. Fluid Mech.}\ }\textbf {\bibinfo {volume}
  {189}},\ \bibinfo {pages} {165} (\bibinfo {year} {1988})}\BibitemShut
  {NoStop}%
\bibitem [{\citenamefont {Stokes}\ \emph {et~al.}(1986)\citenamefont {Stokes},
  \citenamefont {Weitz}, \citenamefont {Gollub}, \citenamefont {Dougherty},
  \citenamefont {Robbins}, \citenamefont {Chaikin},\ and\ \citenamefont
  {Lindsay}}]{stokes-prl-1986}%
  \BibitemOpen
  \bibfield  {author} {\bibinfo {author} {\bibfnamefont {J.~P.}\ \bibnamefont
  {Stokes}}, \bibinfo {author} {\bibfnamefont {D.~A.}\ \bibnamefont {Weitz}},
  \bibinfo {author} {\bibfnamefont {J.~P.}\ \bibnamefont {Gollub}}, \bibinfo
  {author} {\bibfnamefont {A.}~\bibnamefont {Dougherty}}, \bibinfo {author}
  {\bibfnamefont {M.~O.}\ \bibnamefont {Robbins}}, \bibinfo {author}
  {\bibfnamefont {P.~M.}\ \bibnamefont {Chaikin}},\ and\ \bibinfo {author}
  {\bibfnamefont {H.~M.}\ \bibnamefont {Lindsay}},\ }\bibfield  {title}
  {\bibinfo {title} {Interfacial stability of immiscible displacement in a
  porous medium},\ }\href@noop {} {\bibfield  {journal} {\bibinfo  {journal}
  {Phys. Rev. Lett.}\ }\textbf {\bibinfo {volume} {57}},\ \bibinfo {pages}
  {1718} (\bibinfo {year} {1986})}\BibitemShut {NoStop}%
\bibitem [{\citenamefont {Cieplak}\ and\ \citenamefont
  {Robbins}(1988)}]{cieplak-prl-1988}%
  \BibitemOpen
  \bibfield  {author} {\bibinfo {author} {\bibfnamefont {M.}~\bibnamefont
  {Cieplak}}\ and\ \bibinfo {author} {\bibfnamefont {M.~O.}\ \bibnamefont
  {Robbins}},\ }\bibfield  {title} {\bibinfo {title} {Dynamical transition in
  quasistatic fluid invasion in porous media},\ }\href@noop {} {\bibfield
  {journal} {\bibinfo  {journal} {Phys. Rev. Lett.}\ }\textbf {\bibinfo
  {volume} {60}},\ \bibinfo {pages} {2042} (\bibinfo {year}
  {1988})}\BibitemShut {NoStop}%
\bibitem [{\citenamefont {Zhao}\ \emph {et~al.}(2016)\citenamefont {Zhao},
  \citenamefont {MacMinn},\ and\ \citenamefont {Juanes}}]{zhao-pnas-2016}%
  \BibitemOpen
  \bibfield  {author} {\bibinfo {author} {\bibfnamefont {B.}~\bibnamefont
  {Zhao}}, \bibinfo {author} {\bibfnamefont {C.~W.}\ \bibnamefont {MacMinn}},\
  and\ \bibinfo {author} {\bibfnamefont {R.}~\bibnamefont {Juanes}},\
  }\bibfield  {title} {\bibinfo {title} {Wettability control on multiphase flow
  in patterned microfluidics},\ }\href@noop {} {\bibfield  {journal} {\bibinfo
  {journal} {Proc. Natl. Acad. Sci. USA}\ }\textbf {\bibinfo {volume} {113}},\
  \bibinfo {pages} {10251} (\bibinfo {year} {2016})}\BibitemShut {NoStop}%
\bibitem [{\citenamefont {Primkulov}\ \emph {et~al.}(2021)\citenamefont
  {Primkulov}, \citenamefont {Pahlavan}, \citenamefont {Fu}, \citenamefont
  {Zhao}, \citenamefont {MacMinn},\ and\ \citenamefont
  {Juanes}}]{primkulov-jfm-2021}%
  \BibitemOpen
  \bibfield  {author} {\bibinfo {author} {\bibfnamefont {B.~K.}\ \bibnamefont
  {Primkulov}}, \bibinfo {author} {\bibfnamefont {A.~A.}\ \bibnamefont
  {Pahlavan}}, \bibinfo {author} {\bibfnamefont {X.}~\bibnamefont {Fu}},
  \bibinfo {author} {\bibfnamefont {B.}~\bibnamefont {Zhao}}, \bibinfo {author}
  {\bibfnamefont {C.~W.}\ \bibnamefont {MacMinn}},\ and\ \bibinfo {author}
  {\bibfnamefont {R.}~\bibnamefont {Juanes}},\ }\bibfield  {title} {\bibinfo
  {title} {Wettability and lenormand's diagram},\ }\href@noop {} {\bibfield
  {journal} {\bibinfo  {journal} {J. Fluid Mech.}\ }\textbf {\bibinfo {volume}
  {923}},\ \bibinfo {pages} {A34} (\bibinfo {year} {2021})}\BibitemShut
  {NoStop}%
\bibitem [{\citenamefont {Powers}\ \emph {et~al.}(1996)\citenamefont {Powers},
  \citenamefont {Anckner},\ and\ \citenamefont {Seacord}}]{powers-jee-1996}%
  \BibitemOpen
  \bibfield  {author} {\bibinfo {author} {\bibfnamefont {S.~E.}\ \bibnamefont
  {Powers}}, \bibinfo {author} {\bibfnamefont {W.~H.}\ \bibnamefont
  {Anckner}},\ and\ \bibinfo {author} {\bibfnamefont {T.~F.}\ \bibnamefont
  {Seacord}},\ }\bibfield  {title} {\bibinfo {title} {{Wettability of
  NAPL-contaminated sands}},\ }\href@noop {} {\bibfield  {journal} {\bibinfo
  {journal} {J. Environ. Eng.}\ }\textbf {\bibinfo {volume} {122}},\ \bibinfo
  {pages} {889} (\bibinfo {year} {1996})}\BibitemShut {NoStop}%
\bibitem [{\citenamefont {Sinha}\ and\ \citenamefont
  {Wang}(2008)}]{sinha-ces-2008}%
  \BibitemOpen
  \bibfield  {author} {\bibinfo {author} {\bibfnamefont {P.~K.}\ \bibnamefont
  {Sinha}}\ and\ \bibinfo {author} {\bibfnamefont {C.-Y.}\ \bibnamefont
  {Wang}},\ }\bibfield  {title} {\bibinfo {title} {Liquid water transport in a
  mixed-wet gas diffusion layer of polymer electrolyte fuel cell},\ }\href@noop
  {} {\bibfield  {journal} {\bibinfo  {journal} {Chem. Eng. Sci.}\ }\textbf
  {\bibinfo {volume} {63}},\ \bibinfo {pages} {1081} (\bibinfo {year}
  {2008})}\BibitemShut {NoStop}%
\bibitem [{\citenamefont {Salathiel}(1973)}]{Salathiel1973}%
  \BibitemOpen
  \bibfield  {author} {\bibinfo {author} {\bibfnamefont {R.~A.}\ \bibnamefont
  {Salathiel}},\ }\bibfield  {title} {\bibinfo {title} {Oil recovery by surface
  film drainage in mixed-wettability rocks},\ }\href@noop {} {\bibfield
  {journal} {\bibinfo  {journal} {J. Pet. Technol.}\ }\textbf {\bibinfo
  {volume} {25}},\ \bibinfo {pages} {1216} (\bibinfo {year}
  {1973})}\BibitemShut {NoStop}%
\bibitem [{\citenamefont {Kovscek}\ \emph {et~al.}(1993)\citenamefont
  {Kovscek}, \citenamefont {Wong},\ and\ \citenamefont
  {Radke}}]{kovscek-aiche-1993}%
  \BibitemOpen
  \bibfield  {author} {\bibinfo {author} {\bibfnamefont {A.~R.}\ \bibnamefont
  {Kovscek}}, \bibinfo {author} {\bibfnamefont {H.}~\bibnamefont {Wong}},\ and\
  \bibinfo {author} {\bibfnamefont {C.~J.}\ \bibnamefont {Radke}},\ }\bibfield
  {title} {\bibinfo {title} {A pore-level scenario for the development of mixed
  wettability in oil reservoirs},\ }\href@noop {} {\bibfield  {journal}
  {\bibinfo  {journal} {AIChE J.}\ }\textbf {\bibinfo {volume} {39}},\ \bibinfo
  {pages} {1072} (\bibinfo {year} {1993})}\BibitemShut {NoStop}%
\bibitem [{\citenamefont {AlRatrout}\ \emph {et~al.}(2018)\citenamefont
  {AlRatrout}, \citenamefont {Blunt},\ and\ \citenamefont
  {Bijeljic}}]{AlRatrout2018}%
  \BibitemOpen
  \bibfield  {author} {\bibinfo {author} {\bibfnamefont {A.}~\bibnamefont
  {AlRatrout}}, \bibinfo {author} {\bibfnamefont {M.~J.}\ \bibnamefont
  {Blunt}},\ and\ \bibinfo {author} {\bibfnamefont {B.}~\bibnamefont
  {Bijeljic}},\ }\bibfield  {title} {\bibinfo {title} {Wettability in complex
  porous materials, the mixed-wet state, and its relationship to surface
  roughness},\ }\href@noop {} {\bibfield  {journal} {\bibinfo  {journal} {Proc.
  Natl. Acad. Sci. USA}\ }\textbf {\bibinfo {volume} {115}},\ \bibinfo {pages}
  {8901} (\bibinfo {year} {2018})}\BibitemShut {NoStop}%
\bibitem [{\citenamefont {Blunt}\ \emph {et~al.}(2019)\citenamefont {Blunt},
  \citenamefont {Lin}, \citenamefont {Akai},\ and\ \citenamefont
  {Bijeljic}}]{blunt-jcis-2019}%
  \BibitemOpen
  \bibfield  {author} {\bibinfo {author} {\bibfnamefont {M.~J.}\ \bibnamefont
  {Blunt}}, \bibinfo {author} {\bibfnamefont {Q.}~\bibnamefont {Lin}}, \bibinfo
  {author} {\bibfnamefont {T.}~\bibnamefont {Akai}},\ and\ \bibinfo {author}
  {\bibfnamefont {B.}~\bibnamefont {Bijeljic}},\ }\bibfield  {title} {\bibinfo
  {title} {A thermodynamically consistent characterization of wettability in
  porous media using high-resolution imaging},\ }\href@noop {} {\bibfield
  {journal} {\bibinfo  {journal} {J. Colloid Interface Sci.}\ }\textbf
  {\bibinfo {volume} {552}},\ \bibinfo {pages} {59} (\bibinfo {year}
  {2019})}\BibitemShut {NoStop}%
\bibitem [{\citenamefont {Lin}\ \emph {et~al.}(2019)\citenamefont {Lin},
  \citenamefont {Bijeljic}, \citenamefont {Berg}, \citenamefont {Pini},
  \citenamefont {Blunt},\ and\ \citenamefont {Krevor}}]{lin-pre-2019}%
  \BibitemOpen
  \bibfield  {author} {\bibinfo {author} {\bibfnamefont {Q.}~\bibnamefont
  {Lin}}, \bibinfo {author} {\bibfnamefont {B.}~\bibnamefont {Bijeljic}},
  \bibinfo {author} {\bibfnamefont {S.}~\bibnamefont {Berg}}, \bibinfo {author}
  {\bibfnamefont {R.}~\bibnamefont {Pini}}, \bibinfo {author} {\bibfnamefont
  {M.~J.}\ \bibnamefont {Blunt}},\ and\ \bibinfo {author} {\bibfnamefont
  {S.}~\bibnamefont {Krevor}},\ }\bibfield  {title} {\bibinfo {title} {{Minimal
  surfaces in porous media: Pore-scale imaging of multiphase flow in an
  altered-wettability Bentheimer sandstone}},\ }\href@noop {} {\bibfield
  {journal} {\bibinfo  {journal} {Phys. Rev. E}\ }\textbf {\bibinfo {volume}
  {99}},\ \bibinfo {pages} {063105} (\bibinfo {year} {2019})}\BibitemShut
  {NoStop}%
\bibitem [{\citenamefont {Alhammadi}\ \emph {et~al.}(2020)\citenamefont
  {Alhammadi}, \citenamefont {Gao}, \citenamefont {Akai}, \citenamefont
  {Blunt},\ and\ \citenamefont {Bijeljic}}]{alhammadi-fuel-2020}%
  \BibitemOpen
  \bibfield  {author} {\bibinfo {author} {\bibfnamefont {A.~M.}\ \bibnamefont
  {Alhammadi}}, \bibinfo {author} {\bibfnamefont {Y.}~\bibnamefont {Gao}},
  \bibinfo {author} {\bibfnamefont {T.}~\bibnamefont {Akai}}, \bibinfo {author}
  {\bibfnamefont {M.~J.}\ \bibnamefont {Blunt}},\ and\ \bibinfo {author}
  {\bibfnamefont {B.}~\bibnamefont {Bijeljic}},\ }\bibfield  {title} {\bibinfo
  {title} {{Pore-scale X-ray imaging with measurement of relative permeability,
  capillary pressure and oil recovery in a mixed-wet micro-porous carbonate
  reservoir rock}},\ }\href@noop {} {\bibfield  {journal} {\bibinfo  {journal}
  {Fuel}\ }\textbf {\bibinfo {volume} {268}},\ \bibinfo {pages} {117018}
  (\bibinfo {year} {2020})}\BibitemShut {NoStop}%
\bibitem [{\citenamefont {Scanziani}\ \emph {et~al.}(2020)\citenamefont
  {Scanziani}, \citenamefont {Lin}, \citenamefont {Alhosani}, \citenamefont
  {Blunt},\ and\ \citenamefont {Bijeljic}}]{scanziani-royal-2020}%
  \BibitemOpen
  \bibfield  {author} {\bibinfo {author} {\bibfnamefont {A.}~\bibnamefont
  {Scanziani}}, \bibinfo {author} {\bibfnamefont {Q.}~\bibnamefont {Lin}},
  \bibinfo {author} {\bibfnamefont {A.}~\bibnamefont {Alhosani}}, \bibinfo
  {author} {\bibfnamefont {M.~J.}\ \bibnamefont {Blunt}},\ and\ \bibinfo
  {author} {\bibfnamefont {B.}~\bibnamefont {Bijeljic}},\ }\bibfield  {title}
  {\bibinfo {title} {Dynamics of displacement in mixed-wet porous media},\
  }\href@noop {} {\bibfield  {journal} {\bibinfo  {journal} {Proc. R. Soc.
  Lond. A}\ }\textbf {\bibinfo {volume} {476}},\ \bibinfo {pages} {20200040}
  (\bibinfo {year} {2020})}\BibitemShut {NoStop}%
\bibitem [{\citenamefont {Al-Menhali}\ and\ \citenamefont
  {Krevor}(2016)}]{almenhali-krevor-est-2016}%
  \BibitemOpen
  \bibfield  {author} {\bibinfo {author} {\bibfnamefont {A.~S.}\ \bibnamefont
  {Al-Menhali}}\ and\ \bibinfo {author} {\bibfnamefont {S.}~\bibnamefont
  {Krevor}},\ }\bibfield  {title} {\bibinfo {title} {Capillary trapping of
  {CO}$_2$ in oil reservoirs: Observations in a mixed-wet carbonate rock},\
  }\href@noop {} {\bibfield  {journal} {\bibinfo  {journal} {Environ. Sci.
  Technol.}\ }\textbf {\bibinfo {volume} {50}},\ \bibinfo {pages} {2727}
  (\bibinfo {year} {2016})}\BibitemShut {NoStop}%
\bibitem [{\citenamefont {Al-Menhali}\ \emph {et~al.}(2016)\citenamefont
  {Al-Menhali}, \citenamefont {Menke}, \citenamefont {Blunt},\ and\
  \citenamefont {Krevor}}]{almenhali-est-2016}%
  \BibitemOpen
  \bibfield  {author} {\bibinfo {author} {\bibfnamefont {A.~S.}\ \bibnamefont
  {Al-Menhali}}, \bibinfo {author} {\bibfnamefont {H.~P.}\ \bibnamefont
  {Menke}}, \bibinfo {author} {\bibfnamefont {M.~J.}\ \bibnamefont {Blunt}},\
  and\ \bibinfo {author} {\bibfnamefont {S.~C.}\ \bibnamefont {Krevor}},\
  }\bibfield  {title} {\bibinfo {title} {Pore scale observations of trapped
  {CO}$_2$ in mixed-wet carbonate rock: {A}pplications to storage in oil
  fields},\ }\href@noop {} {\bibfield  {journal} {\bibinfo  {journal} {Environ.
  Sci. Technol.}\ }\textbf {\bibinfo {volume} {50}},\ \bibinfo {pages} {10282}
  (\bibinfo {year} {2016})}\BibitemShut {NoStop}%
\bibitem [{\citenamefont {Zou}\ \emph {et~al.}(2018)\citenamefont {Zou},
  \citenamefont {Armstrong}, \citenamefont {Arns}, \citenamefont {Arns},\ and\
  \citenamefont {Hussain}}]{zou-wrr-2018}%
  \BibitemOpen
  \bibfield  {author} {\bibinfo {author} {\bibfnamefont {S.}~\bibnamefont
  {Zou}}, \bibinfo {author} {\bibfnamefont {R.~T.}\ \bibnamefont {Armstrong}},
  \bibinfo {author} {\bibfnamefont {J.-Y.}\ \bibnamefont {Arns}}, \bibinfo
  {author} {\bibfnamefont {C.~H.}\ \bibnamefont {Arns}},\ and\ \bibinfo
  {author} {\bibfnamefont {F.}~\bibnamefont {Hussain}},\ }\bibfield  {title}
  {\bibinfo {title} {Experimental and theoretical evidence for increased
  ganglion dynamics during fractional flow in mixed-wet porous media},\
  }\href@noop {} {\bibfield  {journal} {\bibinfo  {journal} {Water Resour.
  Res.}\ }\textbf {\bibinfo {volume} {54}},\ \bibinfo {pages} {3277} (\bibinfo
  {year} {2018})}\BibitemShut {NoStop}%
\bibitem [{\citenamefont {R{\"u}cker}\ \emph {et~al.}(2019)\citenamefont
  {R{\"u}cker}, \citenamefont {Bartels}, \citenamefont {Singh}, \citenamefont
  {Brussee}, \citenamefont {Coorn}, \citenamefont {van~der Linde},
  \citenamefont {Bonnin}, \citenamefont {Ott}, \citenamefont {Hassanizadeh},
  \citenamefont {Blunt}, \citenamefont {Mahani}, \citenamefont {Georgiadis},\
  and\ \citenamefont {Berg}}]{rucker-grl-2019}%
  \BibitemOpen
  \bibfield  {author} {\bibinfo {author} {\bibfnamefont {M.}~\bibnamefont
  {R{\"u}cker}}, \bibinfo {author} {\bibfnamefont {W.~B.}\ \bibnamefont
  {Bartels}}, \bibinfo {author} {\bibfnamefont {K.}~\bibnamefont {Singh}},
  \bibinfo {author} {\bibfnamefont {N.}~\bibnamefont {Brussee}}, \bibinfo
  {author} {\bibfnamefont {A.}~\bibnamefont {Coorn}}, \bibinfo {author}
  {\bibfnamefont {H.~A.}\ \bibnamefont {van~der Linde}}, \bibinfo {author}
  {\bibfnamefont {A.}~\bibnamefont {Bonnin}}, \bibinfo {author} {\bibfnamefont
  {H.}~\bibnamefont {Ott}}, \bibinfo {author} {\bibfnamefont {S.~M.}\
  \bibnamefont {Hassanizadeh}}, \bibinfo {author} {\bibfnamefont {M.~J.}\
  \bibnamefont {Blunt}}, \bibinfo {author} {\bibfnamefont {H.}~\bibnamefont
  {Mahani}}, \bibinfo {author} {\bibfnamefont {A.}~\bibnamefont {Georgiadis}},\
  and\ \bibinfo {author} {\bibfnamefont {S.}~\bibnamefont {Berg}},\ }\bibfield
  {title} {\bibinfo {title} {The effect of mixed wettability on pore-scale flow
  regimes based on a flooding experiment in ketton limestone},\ }\href@noop {}
  {\bibfield  {journal} {\bibinfo  {journal} {Geophys. Res. Lett.}\ }\textbf
  {\bibinfo {volume} {46}},\ \bibinfo {pages} {3225} (\bibinfo {year}
  {2019})}\BibitemShut {NoStop}%
\bibitem [{\citenamefont {Levach{\'e}}\ \emph {et~al.}(2012)\citenamefont
  {Levach{\'e}}, \citenamefont {Azioune}, \citenamefont {Bourrel},
  \citenamefont {Studer},\ and\ \citenamefont
  {Bartolo}}]{levache-labchip-2012}%
  \BibitemOpen
  \bibfield  {author} {\bibinfo {author} {\bibfnamefont {B.}~\bibnamefont
  {Levach{\'e}}}, \bibinfo {author} {\bibfnamefont {A.}~\bibnamefont
  {Azioune}}, \bibinfo {author} {\bibfnamefont {M.}~\bibnamefont {Bourrel}},
  \bibinfo {author} {\bibfnamefont {V.}~\bibnamefont {Studer}},\ and\ \bibinfo
  {author} {\bibfnamefont {D.}~\bibnamefont {Bartolo}},\ }\bibfield  {title}
  {\bibinfo {title} {Engineering the surface properties of microfluidic
  stickers},\ }\href@noop {} {\bibfield  {journal} {\bibinfo  {journal} {Lab
  Chip}\ }\textbf {\bibinfo {volume} {12}},\ \bibinfo {pages} {3028} (\bibinfo
  {year} {2012})}\BibitemShut {NoStop}%
\bibitem [{\citenamefont {Levach{\'e}}\ and\ \citenamefont
  {Bartolo}(2014)}]{levache-prl-2014}%
  \BibitemOpen
  \bibfield  {author} {\bibinfo {author} {\bibfnamefont {B.}~\bibnamefont
  {Levach{\'e}}}\ and\ \bibinfo {author} {\bibfnamefont {D.}~\bibnamefont
  {Bartolo}},\ }\bibfield  {title} {\bibinfo {title} {Revisiting the
  {S}affman-{T}aylor experiment: {I}mbibition patterns and liquid-entrainment
  transitions},\ }\href@noop {} {\bibfield  {journal} {\bibinfo  {journal}
  {Phys. Rev. Lett.}\ }\textbf {\bibinfo {volume} {113}},\ \bibinfo {pages}
  {044501} (\bibinfo {year} {2014})}\BibitemShut {NoStop}%
\bibitem [{sup()}]{supplement}%
  \BibitemOpen
  \bibinfo {note} {See supplemental material.}\BibitemShut {Stop}%
\bibitem [{\citenamefont {Armstrong}\ and\ \citenamefont
  {Wildenschild}(2012)}]{Armstrong2012}%
  \BibitemOpen
  \bibfield  {author} {\bibinfo {author} {\bibfnamefont {R.~T.}\ \bibnamefont
  {Armstrong}}\ and\ \bibinfo {author} {\bibfnamefont {D.}~\bibnamefont
  {Wildenschild}},\ }\bibfield  {title} {\bibinfo {title} {Microbial enhanced
  oil recovery in fractional-wet systems: A pore-scale investigation},\
  }\href@noop {} {\bibfield  {journal} {\bibinfo  {journal} {Transport Porous
  Med.}\ }\textbf {\bibinfo {volume} {92}},\ \bibinfo {pages} {819} (\bibinfo
  {year} {2012})}\BibitemShut {NoStop}%
\bibitem [{\citenamefont {Brakke}(1992)}]{brakke1992surface}%
  \BibitemOpen
  \bibfield  {author} {\bibinfo {author} {\bibfnamefont {K.~A.}\ \bibnamefont
  {Brakke}},\ }\bibfield  {title} {\bibinfo {title} {The surface evolver},\
  }\href@noop {} {\bibfield  {journal} {\bibinfo  {journal} {Experimental
  mathematics}\ }\textbf {\bibinfo {volume} {1}},\ \bibinfo {pages} {141}
  (\bibinfo {year} {1992})}\BibitemShut {NoStop}%
\bibitem [{\citenamefont {Zhao}\ \emph {et~al.}(2019)\citenamefont {Zhao},
  \citenamefont {MacMinn}, \citenamefont {Primkulov}, \citenamefont {Chen},
  \citenamefont {Valocchi}, \citenamefont {Zhao}, \citenamefont {Kang},
  \citenamefont {Bruning}, \citenamefont {McClure}, \citenamefont {Miller}
  \emph {et~al.}}]{zhao-pnas-2019}%
  \BibitemOpen
  \bibfield  {author} {\bibinfo {author} {\bibfnamefont {B.}~\bibnamefont
  {Zhao}}, \bibinfo {author} {\bibfnamefont {C.~W.}\ \bibnamefont {MacMinn}},
  \bibinfo {author} {\bibfnamefont {B.~K.}\ \bibnamefont {Primkulov}}, \bibinfo
  {author} {\bibfnamefont {Y.}~\bibnamefont {Chen}}, \bibinfo {author}
  {\bibfnamefont {A.~J.}\ \bibnamefont {Valocchi}}, \bibinfo {author}
  {\bibfnamefont {J.}~\bibnamefont {Zhao}}, \bibinfo {author} {\bibfnamefont
  {Q.}~\bibnamefont {Kang}}, \bibinfo {author} {\bibfnamefont {K.}~\bibnamefont
  {Bruning}}, \bibinfo {author} {\bibfnamefont {J.~E.}\ \bibnamefont
  {McClure}}, \bibinfo {author} {\bibfnamefont {C.~T.}\ \bibnamefont {Miller}},
  \emph {et~al.},\ }\bibfield  {title} {\bibinfo {title} {Comprehensive
  comparison of pore-scale models for multiphase flow in porous media},\
  }\href@noop {} {\bibfield  {journal} {\bibinfo  {journal} {Proc. Natl. Acad.
  Sci. U.S.A.}\ }\textbf {\bibinfo {volume} {116}},\ \bibinfo {pages} {13799}
  (\bibinfo {year} {2019})}\BibitemShut {NoStop}%
\bibitem [{\citenamefont {Blunt}(2001)}]{blunt-cocis-2001}%
  \BibitemOpen
  \bibfield  {author} {\bibinfo {author} {\bibfnamefont {M.~J.}\ \bibnamefont
  {Blunt}},\ }\bibfield  {title} {\bibinfo {title} {Flow in porous media ---
  pore-network models and multiphase flow},\ }\href@noop {} {\bibfield
  {journal} {\bibinfo  {journal} {Curr. Opin. Colloid Interface Sci.}\ }\textbf
  {\bibinfo {volume} {6}} (\bibinfo {year} {2001})}\BibitemShut {NoStop}%
\bibitem [{\citenamefont {Cieplak}\ and\ \citenamefont
  {Robbins}(1990)}]{cieplakrobbins90}%
  \BibitemOpen
  \bibfield  {author} {\bibinfo {author} {\bibfnamefont {M.}~\bibnamefont
  {Cieplak}}\ and\ \bibinfo {author} {\bibfnamefont {M.~O.}\ \bibnamefont
  {Robbins}},\ }\bibfield  {title} {\bibinfo {title} {Influence of contact
  angle on quasistatic fluid invasion of porous media},\ }\href@noop {}
  {\bibfield  {journal} {\bibinfo  {journal} {Phys. Rev. B}\ }\textbf {\bibinfo
  {volume} {41}},\ \bibinfo {pages} {11508} (\bibinfo {year}
  {1990})}\BibitemShut {NoStop}%
\bibitem [{\citenamefont {Holtzman}\ and\ \citenamefont
  {Segre}(2015)}]{holtzman-prl-2015}%
  \BibitemOpen
  \bibfield  {author} {\bibinfo {author} {\bibfnamefont {R.}~\bibnamefont
  {Holtzman}}\ and\ \bibinfo {author} {\bibfnamefont {E.}~\bibnamefont
  {Segre}},\ }\bibfield  {title} {\bibinfo {title} {Wettability stabilizes
  fluid invasion into porous media via nonlocal, cooperative pore filling},\
  }\href@noop {} {\bibfield  {journal} {\bibinfo  {journal} {Phys. Rev. Lett.}\
  }\textbf {\bibinfo {volume} {115}},\ \bibinfo {pages} {164501} (\bibinfo
  {year} {2015})}\BibitemShut {NoStop}%
\bibitem [{\citenamefont {Primkulov}\ \emph {et~al.}(2019)\citenamefont
  {Primkulov}, \citenamefont {Pahlavan}, \citenamefont {Fu}, \citenamefont
  {Zhao}, \citenamefont {MacMinn},\ and\ \citenamefont
  {Juanes}}]{Primkulov2019}%
  \BibitemOpen
  \bibfield  {author} {\bibinfo {author} {\bibfnamefont {B.~K.}\ \bibnamefont
  {Primkulov}}, \bibinfo {author} {\bibfnamefont {A.~A.}\ \bibnamefont
  {Pahlavan}}, \bibinfo {author} {\bibfnamefont {X.}~\bibnamefont {Fu}},
  \bibinfo {author} {\bibfnamefont {B.}~\bibnamefont {Zhao}}, \bibinfo {author}
  {\bibfnamefont {C.~W.}\ \bibnamefont {MacMinn}},\ and\ \bibinfo {author}
  {\bibfnamefont {R.}~\bibnamefont {Juanes}},\ }\bibfield  {title} {\bibinfo
  {title} {Signatures of fluid-fluid displacement in porous media: Wettability,
  patterns and pressures},\ }\href@noop {} {\bibfield  {journal} {\bibinfo
  {journal} {Journal of Fluid Mechanics}\ }\textbf {\bibinfo {volume} {875}}
  (\bibinfo {year} {2019})}\BibitemShut {NoStop}%
\bibitem [{\citenamefont {Berg}\ \emph {et~al.}(2013)\citenamefont {Berg},
  \citenamefont {Ott}, \citenamefont {Klapp}, \citenamefont {Schwing},
  \citenamefont {Neiteler}, \citenamefont {Brussee}, \citenamefont {Makurat},
  \citenamefont {Leu}, \citenamefont {Enzmann}, \citenamefont {Schwarz},
  \citenamefont {Kersten}, \citenamefont {Irvine},\ and\ \citenamefont
  {Stampanoni}}]{berg-pnas-2013}%
  \BibitemOpen
  \bibfield  {author} {\bibinfo {author} {\bibfnamefont {S.}~\bibnamefont
  {Berg}}, \bibinfo {author} {\bibfnamefont {H.}~\bibnamefont {Ott}}, \bibinfo
  {author} {\bibfnamefont {S.~A.}\ \bibnamefont {Klapp}}, \bibinfo {author}
  {\bibfnamefont {A.}~\bibnamefont {Schwing}}, \bibinfo {author} {\bibfnamefont
  {R.}~\bibnamefont {Neiteler}}, \bibinfo {author} {\bibfnamefont
  {N.}~\bibnamefont {Brussee}}, \bibinfo {author} {\bibfnamefont
  {A.}~\bibnamefont {Makurat}}, \bibinfo {author} {\bibfnamefont
  {L.}~\bibnamefont {Leu}}, \bibinfo {author} {\bibfnamefont {F.}~\bibnamefont
  {Enzmann}}, \bibinfo {author} {\bibfnamefont {J.}~\bibnamefont {Schwarz}},
  \bibinfo {author} {\bibfnamefont {M.}~\bibnamefont {Kersten}}, \bibinfo
  {author} {\bibfnamefont {S.}~\bibnamefont {Irvine}},\ and\ \bibinfo {author}
  {\bibfnamefont {M.}~\bibnamefont {Stampanoni}},\ }\bibfield  {title}
  {\bibinfo {title} {Real-time 3d imaging of haines jumps in porous media
  flow},\ }\href@noop {} {\bibfield  {journal} {\bibinfo  {journal} {Proc.
  Natl. Acad. Sci. USA}\ }\textbf {\bibinfo {volume} {110}},\ \bibinfo {pages}
  {3755} (\bibinfo {year} {2013})}\BibitemShut {NoStop}%
\bibitem [{\citenamefont {Haines}(20)}]{haines-jas-1930}%
  \BibitemOpen
  \bibfield  {author} {\bibinfo {author} {\bibfnamefont {W.~B.}\ \bibnamefont
  {Haines}},\ }\bibfield  {title} {\bibinfo {title} {Studies in the physical
  properties of soil. v. the hysteresis effect in capillary properties, and the
  modes of moisture distribution associated therewith},\ }\href@noop {}
  {\bibfield  {journal} {\bibinfo  {journal} {J. Agric. Sci.}\ } (\bibinfo
  {year} {20})}\BibitemShut {NoStop}%
\end{thebibliography}%


%

\end{document}